\documentclass[prl,twocolumn,showpacs,10pt,floatfix]{revtex4}

\usepackage{graphicx}
\usepackage{amsmath}
\usepackage{amssymb}
\usepackage{citesort}

\newcommand{\identity}[1]{#1}  

\newcommand{\<}[1]{\hspace{-0.11111em}#1\hspace{-0.11111em}}

\DeclareRobustCommand{\scrE}{\ensuremath{\mathcal{E}}}
\DeclareRobustCommand{\rootstinline}{\ensuremath{\sqrt{7}/2}}

\DeclareRobustCommand{\nhat}{\ensuremath{\hat{n}}}
\DeclareRobustCommand{\Qhat}{\ensuremath{\hat{Q}}}
\DeclareRobustCommand{\betap}{\ensuremath{\beta_\pi}}  
\DeclareRobustCommand{\betan}{\ensuremath{\beta_\nu}}  

\DeclareRobustCommand{\grp}[1]{\mathrm{#1}}

\DeclareRobustCommand{\grpusixp}{\ensuremath{\grp{U}_{\pi}(6)}}
\DeclareRobustCommand{\grpusixn}{\ensuremath{\grp{U}_{\nu}(6)}}
\DeclareRobustCommand{\grpupn}{\ensuremath{\grp{U}_{\pi\nu}(5)}}
\DeclareRobustCommand{\grpsopn}{\ensuremath{\grp{SO}_{\pi\nu}(6)}}
\DeclareRobustCommand{\grpsupn}{\ensuremath{\grp{SU}_{\pi\nu}(3)}}
\DeclareRobustCommand{\grpsupnbar}{\ensuremath{\overline{\grp{SU}_{\pi\nu}(3)}}}
\DeclareRobustCommand{\grpsupnstar}{\ensuremath{\grp{SU}_{\pi\nu}^*(3)}}
\DeclareRobustCommand{\grpsupnstarbar}{\ensuremath{\overline{\grp{SU}_{\pi\nu}^*(3)}}}

\begin{document}


\title{Phase structure of the two-fluid proton-neutron system}

\author{M. A. Caprio}
\author{F. Iachello}
\affiliation{Center for Theoretical Physics, Sloane Physics Laboratory, 
Yale University, New Haven, Connecticut 06520-8120, USA}
\affiliation{European Centre for Theoretical
Studies in Nuclear Physics and Related Areas,
Strada delle Tabarelle 286, 38050 Villazzano (Trento), Italy}

\date{\today}

\begin{abstract}
The phase structure of a two-fluid bosonic system is investigated.
The proton-neutron interacting boson model (IBM-2) posesses a rich
phase structure involving three control parameters and multiple order
parameters.  The surfaces of quantum phase transition between
spherical, axially-symmetric deformed, and $\grpsupnstar$ triaxial
phases are determined.
\end{abstract}


\maketitle


The phase structure of quantum many-body systems has in recent years
been a subject of great experimental and theoretical interest.  Models
based upon algebraic Hamiltonians have found extensive application to
the spectroscopy of many-body systems, including
nuclei~\cite{iachello1987:ibm} and
molecules~\cite{iachello1995:vibron}.  Applications to hadrons have
also been developed~\cite{bijker1994:algebraic-hadron-nonstrange}.
For certain specific forms of its Hamiltonian, an algebraic model
exhibits dynamical symmetries.  In the classical limit, these
dynamical symmetries correspond to qualitatively distinct ground-state
equilibrium configurations, which constitute the phases of the
system~\cite{gilmore1978:coherent,feng1981:ibm-phase}.  The phase
structure has been studied in detail for algebraic models describing
systems composed of one species of constituent particle (``one-fluid''
systems), in particular the interacting boson
model~(IBM)~\cite{iachello1987:ibm} for nuclei.  While one-fluid
systems are described by a single elementary Lie algebra, usually
$\grp{U}(n)$, multi-fluid systems are described by a coupling of such
Lie algebras,
$\grp{U}_1(n)\otimes\grp{U}_2(n)\otimes\cdots$~\cite{iachello1987:ibm,iachello1995:vibron}.
This more involved algebraic structure naturally leads to a richer
phase structure.

In the present work, the phase structure of a system comprised of two
interacting fluids is investigated.  The proton-neutron interacting
boson model~(IBM-2)~\cite{arima1977:ibm2-shell,iachello1987:ibm}, in
which proton pairs and neutron pairs are treated as distinct
constituents, is considered.  While the one-fluid IBM exhibits three
dynamical symmetries, separated by first and second order phase
transitions~\cite{dieperink1980:ibm-classical,feng1981:ibm-phase}, the
IBM-2 supports four dynamical
symmetries~\cite{vanisacker1986:ibm2-limits,dieperink1982:ibm2-triax},
and thus inherently has a higher-dimensional phase diagram.  Moreover,
the phase structure is found to posess qualitatively new features.
Due to the complexity of the problem, a combination of analytic and
numerical methods have been applied in this analysis.  Preliminary
investigations of the IBM-2 phase structure have been presented in
Refs.~\nocite{proc-icnpls04}\cite{arias2004:ibmpn-icnpls04,caprio2004:ibmpn-icnpls04}.

Before proceeding, let us briefly summarize the IBM-2 Hamiltonian and
the dynamical symmetries it supports.  Operators in the IBM-2 are
constructed from the generators of the group
$\grpusixp\otimes\grpusixn$, realized in terms of the boson creation
operators $s_{\rho,0}^{\dag}$ and $d_{\rho,\mu}^{\dag}$ (where $\rho$
represents $\pi$ or $\nu$, and $\mu$$=$$-2,\ldots,2$) and their
associated annihilation operators, acting on a basis of good boson
numbers $N_\pi$ and $N_\nu$.  A schematic Hamiltonian which retains
the essential features of the model is the $F$-spin invariant
Hamiltonian (\textit{e.g.}, Ref.~\cite{lipas1990:ibm2-fspin})
\begin{equation}
\label{eqnHepsilon}
H=\varepsilon ( \nhat_{d\pi}+\nhat_{d\nu})+
\kappa(\Qhat_\pi^{\chi_\pi}+\Qhat_\nu^{\chi_\nu})\cdot(\Qhat_\pi^{\chi_\pi}+\Qhat_\nu^{\chi_\nu}),
\end{equation}
where $\nhat_{d\rho}\<\equiv d_\rho^{\dag}\<\cdot\tilde{d}_\rho$ and
$\Qhat_\rho^{\chi_\rho}\<\equiv(s_\rho^{\dag}\<\times\tilde{d}_\rho +
d_\rho^{\dag}\<\times\tilde{s}_\rho)^{(2)} +
\chi_\rho(d_\rho^{\dag}\<\times\tilde{d}_\rho)^{(2)}$.  It is
convenient to also introduce ``scalar'' and ``vector'' parameters
$\chi_S$$\equiv$$\frac{1}{2}(\chi_\pi+\chi_\nu)$ and
$\chi_V$$\equiv$$\frac{1}{2}(\chi_\pi-\chi_\nu)$.  Three of the IBM-2
symmetries occur for $\chi_V$$=$$0$ and have direct analogues in the
one-fluid IBM~\cite{vanisacker1986:ibm2-limits}: $\grpupn$
($\kappa\<=0$), for which the geometric interpretation is that of
undeformed proton and neutron fluids, $\grpsopn$ ($\varepsilon\<=0$,
$\chi_\pi\<=\chi_\nu\<=0$), yielding deformed, $\gamma$-unstable
structure, and $\grpsupn$ ($\varepsilon\<=0$,
$\chi_\pi\<=\chi_\nu\<=-\rootstinline$), for which prolate axially
symmetric structure is obtained.  [The complementary case
$\chi_\pi$$=$$\chi_\nu$$=$$+\rootstinline$, giving oblate axially
symmetric structure, is distinguished by the notation $\grpsupnbar$.]
However, a symmetry special to the IBM-2, denoted $\grpsupnstar$, is
obtained for $\varepsilon\<=0$, $\chi_\pi\<=-\rootstinline$, and
$\chi_\nu\<=+\rootstinline$~\cite{dieperink1982:ibm2-triax}.  The
equilibrium configuration consists of a proton fluid with axially
symmetric prolate deformation coupled to a neutron fluid with axially
symmetric oblate deformation, with their symmetry axes orthogonal to
each
other~\cite{dieperink1982:ibm2-triax,leviatan1990:ibm2-modes,ginocchio1992:ibm2-shapes},
yielding an overall composite nuclear shape with triaxial deformation.
To avoid ambiguity, we shall adopt here the notation $\grpsupnstarbar$
for the complementary case $\chi_\pi\<=+\rootstinline$ and
$\chi_\nu\<=-\rootstinline$, for which the proton and neutron
deformations are interchanged.

The classical limit of the IBM-2 is obtained by evaluation of the
expectation value of $H$ for the coherent state
$|N_\pi,\alpha^{(2)}_\pi;N_\nu,\alpha^{(2)}_\nu\rangle$$\propto$ $(
s_{\pi,0}^\dagger+\sum_\mu\alpha^{(2)}_{\pi,\mu} d_{\pi,\mu}^{\dagger}
)^{N_\pi} ( s_{\nu,0}^\dagger+\sum_\mu\alpha^{(2)}_{\nu,\mu}
d_{\nu,\mu}^{\dagger} )^{N_\nu} |0\rangle$.  This yields an energy
surface $\scrE$$\equiv$$\langle H
\rangle$ as a function of the coherent state parameters
$\alpha^{(2)}_{\rho,\mu}$.  The $\alpha^{(2)}_{\rho,\mu}$ are
interpreted geometrically as the quadrupole shape
variables~\cite{bohr1998:v2} for the proton and neutron fluids and are
equivalent to four deformation parameters ($\beta_\pi$, $\gamma_\pi$,
$\beta_\nu$, and $\gamma_\nu$) and six Euler angles ($\theta_{1\pi}$,
$\theta_{2\pi}$, $\theta_{3\pi}$, $\theta_{1\nu}$, $\theta_{2\nu}$,
and $\theta_{3\nu}$).  By rotational invariance, $\scrE$ depends only
upon the \textit{relative} Euler angles $\vartheta_i$ between the
proton and neutron fluid intrinsic systems, not the $\theta_{i\pi}$
and $\theta_{i\nu}$ separately.  Minimization of
$\scrE(\beta_\pi,\gamma_\pi,\beta_\nu,
\gamma_\nu,\vartheta_1,\vartheta_2,\vartheta_3)$ with respect to
its parameters yields the classical equilibrium configuration of the
ground state.  The terms $\langle \nhat_{d\pi}\rangle$, $\langle
\nhat_{d\nu}\rangle$, $\langle \Qhat_\pi \cdot \Qhat_\pi\rangle$, and $\langle
\Qhat_\nu \cdot \Qhat_\nu\rangle$ contributing to $\scrE$ involve only a
single fluid and are thus known from the IBM (see
Ref.~\cite{vanisacker1981:ibm-triax}).  The expectation value $\langle
\Qhat_\pi\cdot \Qhat_\nu\rangle$ of the interaction term is obtained by the
methods of
Refs.~\cite{vanisacker1981:ibm-triax,ginocchio1992:ibm2-shapes} as a
function of all seven possible parameters ($\beta_\pi$, $\gamma_\pi$,
$\beta_\nu$, $\gamma_\nu$, $\vartheta_1$, $\vartheta_2$, and
$\vartheta_3$),
\begin{widetext}
\begin{equation}
\label{equationQpQn}
\langle \Qhat_\pi^{\chi_\pi} \cdot \Qhat_\nu^{\chi_\nu} \rangle 
=
\frac{N_\pi N_\nu}{(1+\beta_\pi^2)(1+\beta_\nu^2)}
[\alpha_\pi^{(2)\,*}+\tilde\alpha_\pi^{(2)}+\chi_\pi
(\alpha_\pi^{(2)\,*}\times\tilde\alpha_\pi^{(2)})^{(2)}]\\
\cdot
[\alpha_\nu^{(2)\,*}+\tilde\alpha_\nu^{(2)}+\chi_\nu(\alpha_\nu^{(2)\,*}\times\tilde\alpha_\nu^{(2)})^{(2)}]
,
\end{equation}
\end{widetext}
where
\begin{multline}
\alpha_{\rho,\mu}^{(2)} =
\beta_\rho\cos\gamma_\rho D^{2\,*}_{\mu0}(\theta_{1\rho},\theta_{2\rho},\theta_{3\rho})\\
+\frac{1}{\sqrt{2}}\beta_\rho\sin\gamma_\rho
\bigl[D^{2\,*}_{\mu2}(\theta_{1\rho},\theta_{2\rho},\theta_{3\rho})+D^{2\,*}_{\mu-2}(\theta_{1\rho},\theta_{2\rho},\theta_{3\rho})\bigr]
\end{multline}
and where the Euler angles may be chosen to be $\theta_{i\pi}\<=0$ and
$\theta_{i\nu}\<=\vartheta_{i}$ by rotational invariance.  The
$\langle \nhat_{d\rho}\rangle$ are linear in $N_\pi$ or $N_\nu$, while the
$\langle \Qhat_\rho\cdot \Qhat_{\rho'}\rangle$ are quadratic. It is thus convenient
to reparametrize the Hamiltonian~(\ref{eqnHepsilon}) as
\begin{equation}
\label{eqnHxi}
H=\frac{1-\xi'}{N} ( \nhat_{d\pi}+\nhat_{d\nu})
-\frac{\xi'}{N^2}(\Qhat_\pi^{\chi_\pi}+\Qhat_\nu^{\chi_\nu})\cdot(\Qhat_\pi^{\chi_\pi}+\Qhat_\nu^{\chi_\nu}),
\end{equation}
where $N\<\equiv N_\pi+N_\nu$, so that the energy function $\scrE$ is
independent of $N$ at fixed ratio $N_\pi/N_\nu$.  This definition also
condenses the full range of possible ratios $\varepsilon/\kappa$ onto
the finite interval $0\<\leq\xi'\<\leq1$.  An overall normalization
parameter for $H$ has been discarded as irrelevant to the structure of
the energy surface.  There are thus three control parameters~---
$\xi'$, $\chi_\pi$, and $\chi_\nu$~--- for this Hamiltonian.

A simple categorization of the possible Euler angle and $\gamma_\rho$
values for the equilibrium configurations for certain IBM-2 energy
surfaces has been presented in Ref.~\cite{ginocchio1992:ibm2-shapes}.
For a class of Hamiltonians including the present one~(\ref{eqnHxi}),
it is found that the global minimum only occurs for vanishing relative
Euler angles, \textit{i.e.}, for aligned proton and neutron intrinsic
frames.  This effectively reduces the number of order parameters for
the system from seven to four~--- $\beta_\pi$, $\gamma_\pi$,
$\beta_\nu$, and $\gamma_\nu$.%
\begin{figure}[b]
\begin{center}
\includegraphics[width=0.9\hsize]{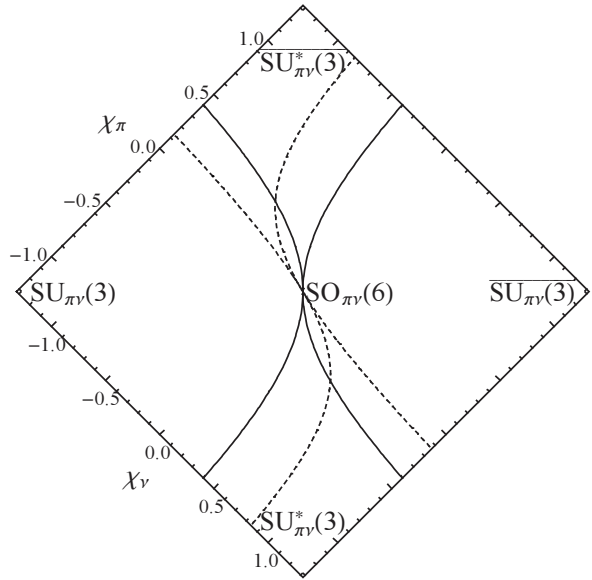}
\end{center}
\vspace{-12pt}
\caption
{Phase diagram for the
$\grp{SU}_{\pi\nu}(3)$-$\grp{SO}_{\pi\nu}(6)$-$\grp{SU}_{\pi\nu}^*(3)$
plane ($\xi'\<=1$) in the parameter space of the Hamiltonian
of~(\ref{eqnHxi}), showing the second-order phase transition
curves~(\ref{eqnboundary}) for the cases $N_\pi/N_\nu\<=1$ (solid) and
$N_\pi/N_\nu\<=4$ (dashed).  The diagram is rotated to allow more
direct comparison with Fig.~\ref{figquadrant}.
\label{figbackplane}
}
\end{figure}

The orders of phase transitions are, in the present study, determined
according to the Ehrenfest classification: a phase transition is first
order if the first derivative of the system's energy is discontinuous
with respect to the control parameter being varied, second order if
the second derivative is discontinuous,
\textit{etc.}  Where the system's energy is obtained, as in the present classical
analysis, as the global minimum of an energy function $\scrE$, a first
order transition is usually associated with a discontinuous jump in
the equilibrium coordinates (``order parameters'') between distinct
competing minima.  Second or higher order transitions are associated
instead with a continuous evolution of the equilibrium coordinates, as
when an initially solitary global minimum becomes unstable (posessing
a vanishing second derivative with respect to some coordinate) and
evolves into two or more minima.  It should be noted that, whenever
the order of a phase transition is obtained by numerical analysis,
application of the Ehrenfest criterion is limited by the ability to
numerically resolve sufficiently small discontinuities, especially a
consideration for points of first-order transition very close to a
point of second-order transition.  Moreover, problems with the
classification scheme, not addressed here, may arise at the boundaries
of the parameter space or when the Hamiltonian posesses additional
symmetries.

We begin our analysis with an analytic study of the phase structure of
the Hamiltonian~(\ref{eqnHxi}) for $\xi'\<=1$, which encompasses the
$\grpsopn$, $\grpsupn$, and $\grpsupnstar$ dynamical symmetries (see
Fig.~\ref{figbackplane}).  The global equilibrium in this case is
always deformed ($\beta_\pi\<>0$ and $\beta_\nu\<>0$).  Surrounding
the $\grpsupn$ dynamical symmetry is a region of parameter space in
which the equilibrium deformations are axially symmetric
($\gamma_\pi\<=\gamma_\nu\<=0^\circ$), and a similar region surrounds
the $\grpsupnbar$ dynamical symmetry
($\gamma_\pi\<=\gamma_\nu\<=60^\circ$).  Taking the $\grpsupn$-like
region for specificity, the global minimum occurs for
\begin{equation}
\label{eqnbeta}
\beta_\rho= \Biggl[\Biggl( \frac{\chi_\rho}{\sqrt{14}}\Biggr)^2+1\Biggr]
^{1/2}-\frac{\chi_\rho}{\sqrt{14}}.
\end{equation}
At the boundary of this region, axial equilibrium deformation gives
way to triaxial deformation, with $\gamma_\pi$ and/or $\gamma_\nu$
nonzero.  This transition occurs continuously on the locus of points
at which the minimum given by~(\ref{eqnbeta}) first becomes unstable
with respect to $\gamma$ deformation.  Since $\scrE$ depends upon both
$\gamma_\pi$ and $\gamma_\nu$, instablility occurs when the
\textit{directional} second derivative of $\scrE$ first vanishes along
some ``direction'' in $(\gamma_\pi,\gamma_\nu)$ coordinate space,
which may generally happen before either
$\partial^2\scrE/\partial\gamma_\pi^2$ or
$\partial^2\scrE/\partial\gamma_\nu^2$ vanishes individually.  The
equation describing the boundary curve in $\chi_\pi$ and $\chi_\nu$ is
most compactly expressed in terms of the corresponding equilibrium
values $\beta_\pi$ and $\beta_\nu$ from~(\ref{eqnbeta}) as
\begin{multline}
\label{eqnboundary}
1 =
\frac{
9\frac{N_\pi}{N_\nu}\betap(\betap^2-1)+\betan(2\betap^2-1)(\betap^2+1)
} {2\betap (\betap^2-2)^2}
\\
\times
\frac{
9\frac{N_\nu}{N_\pi}\betan(\betan^2-1)+\betap(2\betan^2-1)(\betan^2+1)
} {2 \betan (\betan^2-2)^2} .
\end{multline}
This curve is shown in Fig.~\ref{figbackplane}.  Along the
$\grpsupn$-$\grpsupnstar$ line ($\chi_\pi\<=-\rootstinline$), for
$N_\pi/N_\nu\<=1$ the transition occurs at $\chi_\nu\<\approx0.4035$,
at which point the global minimum becomes soft with respect to
$\gamma_\nu$ at fixed $\gamma_\pi\<=0^\circ$.

\begin{figure}
\begin{center}
\includegraphics[width=1.0\hsize]{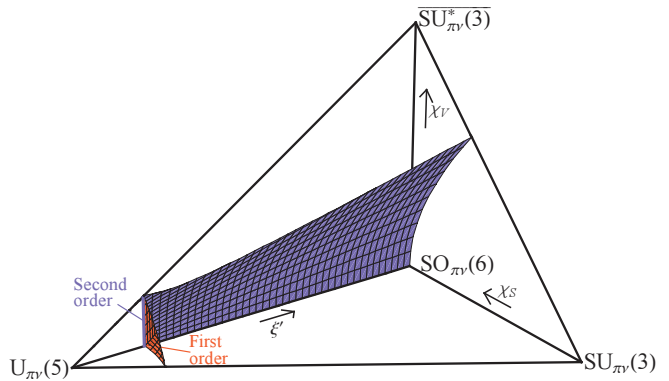}
\end{center}
\vspace{-12pt}
\caption
{Phase diagram of the proton-neutron interacting boson model (IBM-2)
for the Hamiltonian of~(\ref{eqnHxi}), as obtained by numerical
minimization of $\scrE$, for $N_\pi/N_\nu\<=1$.  The surfaces of
first-order (red) and second-order (blue) transition between regions
of undeformed, axially symmetric deformed, and triaxially deformed
equilibria are shown.  Only one ``quadrant'' of parameter space
($0\<\leq\xi'\<\leq 1$, $-\sqrt{7}/2\<\leq\chi_S\<\leq0$, and
$0\<\leq\chi_V\<\leq\sqrt{7}/2$) is included in this plot, since the
others may be obtained by reflection.  The $\chi_S$ and $\chi_V$ axes
are scaled by $\xi'$, so as to converge to a point at the $\grpupn$
limit.
\label{figquadrant}
}
\end{figure}

Returning to the full parameter space for the three-parameter
Hamiltonian of~(\ref{eqnHxi}), limited but useful analytic results can
also be obtained for the transition between undeformed
($\beta_\pi\<=\beta_\nu\<=0$) and deformed structure.  The parameter
space is illustrated in Fig.~\ref{figquadrant}.

For $\chi_V$$=$$0$, \textit{i.e.}, in the ``base'' plane in
Fig.~\ref{figquadrant}, the analysis is closely related to that for
the one-fluid IBM.  The equilibrium configurations all have
$\beta_\pi\<=\beta_\nu(\equiv\hspace{-0.11111em}\beta)$ and are identical to those
obtained for the IBM Hamiltonian $H_\text{IBM}=[(1-\xi')/N] \nhat_d
-(\xi'/N^2)\Qhat^{\chi}\cdot \Qhat^{\chi}$ with $\chi\<=\chi_S$, the phase
structure of which is well
known~\cite{dieperink1980:ibm-classical,feng1981:ibm-phase}.
A second-order transition between undeformed ($\beta\<=0$) and
deformed ($\beta\<\neq0$) structure occurs at the parameter values
$\xi'\<=1/5$ and $\chi\<=0$, for which the minimum in the energy
surface at $\beta\<=0$ is unstable.  This point lies on a trajectory
of first-order transition points, at which a distinct minimum with
nonzero $\beta$ preempts that with $\beta\<=0$ as global minimum.%

A second-order transition from undeformed to deformed structure occurs
when the minimum of $\scrE$ at $\beta_\pi\<=\beta_\nu\<=0$ becomes
unstable with respect to $\beta$ deformation, provided this minimum is
the global minimum (that is, provided it has not been rendered
irrelevant by a prior first-order transition to another, deformed
minimum).  The derivative indicating such $\beta$ softness is the
directional second derivative $\partial^2\scrE/\partial\beta^2$ along
a ray $\beta_\pi\<=u_\pi \beta$ and $\beta_\nu\<= u_\nu \beta$
(\textit{i.e.}, fixed $\beta_\pi/\beta_\nu$) at fixed $\gamma_\pi$ and
$\gamma_\nu$, evaluated at $\beta\<=0$.  This quantity is
\textit{independent} of $\chi_S$ and $\chi_V$ [\textit{e.g.}, for
$N_\pi/N_\nu\<=1$,
$\partial^2\scrE/\partial\beta^2|_{\beta=0}\<=(1-3\xi')(u_\pi^2+u_\nu^2)-4\xi'u_\pi
u_\nu \cos(\gamma_\pi-\gamma_\nu)$].  The minimum at zero deformation
first becomes unstable at $\xi'\<=1/5$, where it is soft against
deformations with $\beta_\pi\<=\beta_\nu$ and
$\gamma_\pi\<=\gamma_\nu(\equiv\hspace{-0.11111em}\gamma)$, for any value of
$\gamma$.  

The curve of first-order phase transition in the plane $\chi_V\<=0$
arises from competition between the undeformed minimum and one with
$\beta_\pi\<=\beta_\nu\<\neq0$ and
$\gamma_\pi\<=\gamma_\nu\<=0^\circ$.  The special ``slice''
$\scrE(\beta_\pi\<=\beta,\beta_\nu\<=\beta,\gamma_\pi\<=0,\gamma_\nu\<=0)$
of the energy function, which includes both these minima, is found for
$N_\pi/N_\nu\<=1$ to be independent of $\chi_V$ at fixed $\xi'$ and
$\chi_S$,
\textit{i.e.}, invariant along any vertical line in
Fig.~\ref{figquadrant}.  Thus, the first-order phase transitions
occuring in the plane $\chi_V\<=0$ at $\xi'\<<1/5$ ``propagate'' out
of this plane.  [To the approximation that
$\beta_\pi\<\approx\beta_\nu$ for the deformed minimum, the
first-order transition surface for $\chi_V\<\neq0$ is obtained by
vertical extension of the one-fluid IBM transition trajectory in
Fig.~\ref{figquadrant}.]  The occurence of a second-order phase
transition at $\xi'\<=1/5$ is thus precluded everywhere except along a
vertical line in parameter space extending through the one-fluid IBM
second-order transition point.  It is verified numerically, as
described below, that the undeformed minimum is indeed global along
this line.  The line $\xi'\<=1/5$ and $\chi_S\<=0$ is thus a locus of
\textit{second-order} transition between zero and nonzero deformation.
These results are readily generalized to arbitrary $N_\pi/N_\nu$, for
which the invariance of $\scrE$ occurs along lines of constant $N_\pi
\chi_\pi + N_\nu \chi_\nu$ and the line of second-order phase transition obeys
$\chi_\pi/\chi_\nu\<=-N_\nu/N_\pi$.

The remainder of the phase diagram is obtained by numerical
minimization of the energy surface with respect to $\beta_\pi$,
$\gamma_\pi$, $\beta_\nu$, and $\gamma_\nu$.  For robust
identification of the global minimum, $\scrE$ is first evaluated at
each point on a fine mesh in these coordinates
($\Delta\beta_\rho\<=0.03$, $\Delta\gamma_\rho\<=2^\circ$), and all
points which are discrete local minima of $\scrE$ relative to the
neighboring mesh points are identified.  The $\beta_\pi$,
$\gamma_\pi$, $\beta_\nu$, and $\gamma_\nu$ values for these minima
are then refined by an iterative method.  The global minimum is
identified from among these.

The phase diagram obtained in this fashion is shown in
Fig.~\ref{figquadrant} for the case $N_\pi/N_\nu\<=1$.  Numerical
investigation of the behavior of $\scrE$ at the transition points
allows the Ehrenfest criterion to be applied, and it appears that the
axial-triaxial transition is everywhere second order.

The IBM-2 phase diagram obtained here provides a framework for
studying the transition between axial and triaxial structure in
nuclei.  Triaxial nuclear deformation might arise from several
different sources: higher-order (cubic, quartic, \textit{etc.})
interactions in an essentially one-fluid
nucleus~\cite{vanisacker1981:ibm-triax}, distinct deformations of the
proton and neutron fluids as discussed here, or the presence of
configurations involving hexadecapole nucleon
pairs~\cite{heyde1983:ibm-g-boson-104ru}.  The present analysis
provides insight into the conditions under which proton-neutron
triaxial deformation may occur and the nature of the transition to
such structure.  Phenomenological studies extending this work should
make use of a more realistic Hamiltonian involving a Majorana
contribution and different strengths for the $\Qhat_\rho\cdot
\Qhat_{\rho'}$ terms~(\textit{e.g.}, Ref.~\cite{lipas1990:ibm2-fspin}).
Such studies will provide guidance for the experimental investigation
of triaxial structure as progressively more neutron rich nuclei become
accessible.

The present analysis may serve as a model for the study of other
multi-fluid bosonic systems.  Within nuclear physics, the
$\grp{U}_\text{core}(6)\otimes\grp{U}_\text{skin}(6)$ description of
core-skin collective modes in neutron rich
nuclei~\cite{warner1997:skin-soft-scissors} can be treated similarly.
In molecular physics, a study of the phase structure is in progress
for the vibron model with two vibronic
species~\cite{iachello1995:vibron}, where it is relevant to coupled
vibronic bending modes in acetylene.  Another area of potential
application of the method is to atomic Bose-Einstein condensates.
Scissors modes, introduced originally within the framework of nuclear
two-rotor models~\cite{loiudice1978:scissors} and the
IBM-2~\cite{iachello-scissors-COMBO}, have been observed for
oscillations of a single-constituent Bose-Einstein condensate relative
to an anisotropic potential~\cite{marago2000:bec-scissors}.
Experiments are planned to produce condensates of two different atomic
species and to study the scissors modes between them.  Triaxial
deformations of the type considered here could then occur, and a
directly analogous analysis could be applied.

The exotic features of the phase diagram considered here, arising from
the presence of multiple control parameters and multiple order
parameters, are likely to be encountered for other multi-fluid systems
as well.  They will require a classification scheme beyond the simple
Ehrenfest or one-parameter Landau models for their proper description.

\begin{acknowledgments}
Discussions with J.~M.~Arias and R.~Bijker are gratefully acknowledged.
This work was supported by the US DOE under grant DE-FG02-91ER-40608.
\end{acknowledgments}

\vfil



\end{document}